\documentclass[conference]{IEEEtran}
\IEEEoverridecommandlockouts
\usepackage{cite}
\usepackage{amsmath,amssymb,amsfonts}
\usepackage{algorithm,algpseudocode}
\usepackage{graphicx}
\usepackage{textcomp}
\usepackage{booktabs}
\usepackage{xcolor}
\def\BibTeX{{\rm B\kern-.05em{\sc i\kern-.025em b}\kern-.08em
    T\kern-.1667em\lower.7ex\hbox{E}\kern-.125emX}}
\begin{document}

\title{HYPERLOCK: In-Memory Hyperdimensional Encryption in Memristor Crossbar Array
}

\author{Jack Cai$^{1}$, Amirali Amirsoleimani$^{2}$, and Roman Genov$^{1}$ \\
$^{1}$Department of Electrical and Computer Engineering, University of Toronto, Toronto, Canada \\
$^{2}$Department of Electrical Engineering and Compute Science, York University, Toronto ON M3J 1P3, Canada \\
Email: $^{1}$jack.cai@mail.utoronto.ca, $^{2}$amirsol@yorku.ca, $^{3}$roman@eecg.utoronto.ca 
}



\maketitle

\begin{abstract}
We present a novel cryptography architecture based on memristor crossbar array, binary hypervectors, and neural network. Utilizing the stochastic and unclonable nature of memristor crossbar and error tolerance of binary hypervectors and neural network, implementation of the algorithm on memristor crossbar simulation is made possible. We demonstrate that with an increasing dimension of the binary hypervectors, the non-idealities in the memristor circuit can be effectively controlled. At the fine level of controlled crossbar non-ideality, noise from memristor circuit can be used to encrypt data while being sufficiently interpretable by neural network for decryption. We applied our algorithm on image cryptography for proof of concept, and to text en/decryption with 100\% decryption accuracy despite crossbar noises. Our work shows the potential and feasibility of using memristor crossbars as an unclonable stochastic encoder unit of cryptography on top of their existing functionality as a vector-matrix multiplication acceleration device.
\end{abstract}


\section{Introduction}
\vspace{-0.5mm}
Memristors are non-volatile, configurable memory devices \cite{Chua1971, Strukov2008}. Their ability to permanently store variable conductance information makes them good candidates to build analog vector matrix multiplications (VMM) crossbar accelerators. The crossbar performs VMM in O(1) and overcomes the von Neumann bottleneck with in-memory computing \cite{Amirsoleimani2020}, which enables many edge computing applications in machine learning such as CNN \cite{Yao2020}, LSTM \cite{Li2019}, and neuromorphic computing \cite{Rahimi2020}. However, despite their high efficiency, low footprint, and low power consumption \cite{Ielmini2018, Sebastian2020, LeGallo2018}, memristor crossbar suffer non-ideality issues such as sneak path current, device variability, stuck conductance, and cycle-to-cycle variability \cite{Yi2016, Ambrogio2018, Du2017, Lv2021}. In some literature, non-ideality and behaviors of memristors are studied and their stochasticity is exploited for hardware security applications \cite{Lv2021} such as physical unclonable functions (PUFs) \cite{Zhang2018, Jiang2018, Nili2018} and chaotic circuits \cite{Yang2015, Wu2016}. Despite algorithms proposed in these studies covering from generating stochastic sequences for hardware verification to public and private key cryptography, none of which have touched on using memristor crossbar's VMM operation to encrypt the data directly. \par
In in-memory hyperdimensional encryption, we investigate the feasibility of using the memristor crossbar's stochastic VMM operation for data encryption. Encrypting data directly with a memristor crossbar poses a challenge for decryption because the cycle to cycle variability can result in inconsistent ciphertext for the same input. On the other hand, such randomness can be beneficial as it prevents the repetitiveness of the encrypted text. 
The in-memory hyperdimensional computing concept proposed by IBM \cite{Karunaratne2020} demonstrates the robustness of binary hypervectors against memristor crossbar non-ideality. Such discovery leads to the inspiration of this work: utilize binary hypervectors encryption to control the impact of noise generated by memristor crossbar, then train a shallow neural network for decryption. We demonstrate in simulation experiments: 1) the proof of concept on image cryptography, and 2) in-text en/decryption to show that at a fine level of noise, the ciphertext is near 100\% decryptable by the neural network while being unique for each pass.
\begin{figure}[t]
    \centering
    \includegraphics[width=\linewidth]{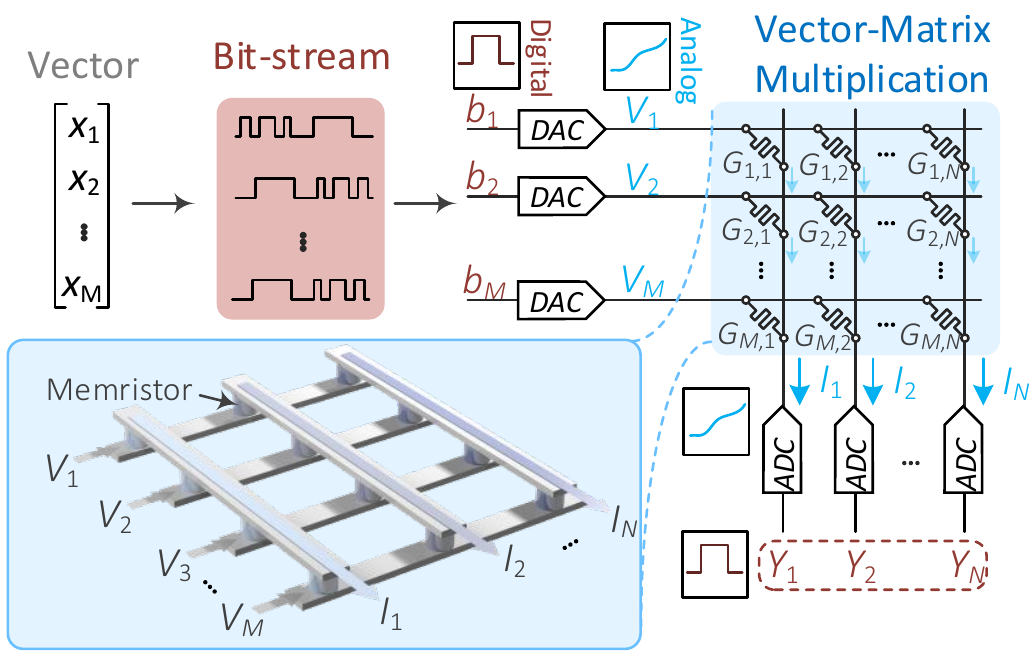}
       \caption{Vector-matrix multiplication (VMM) over the memristor crossbar. 3D crossbar array has been displayed with input voltage and bit-line currents.}
    \label{fig:1}
    \vspace{-6mm}
   \end{figure}
Using a memristor crossbar for this algorithm has a few benefits. First, the cost of VMM operation is low due to in-memory computing. Second, the algorithm can benefit from the intrinsic stochasticity of the memristor crossbar without the need of adding artificial noises. In the end, noises generated by the crossbar provide additional security levels, and the exact properties of the crossbar are unobtainable and unclonable by attackers. The power and time complexity of the memristor crossbar are compared with CMOS digital implementation.
\begin{figure*}[t]
    \centering
    \includegraphics[width=\linewidth]{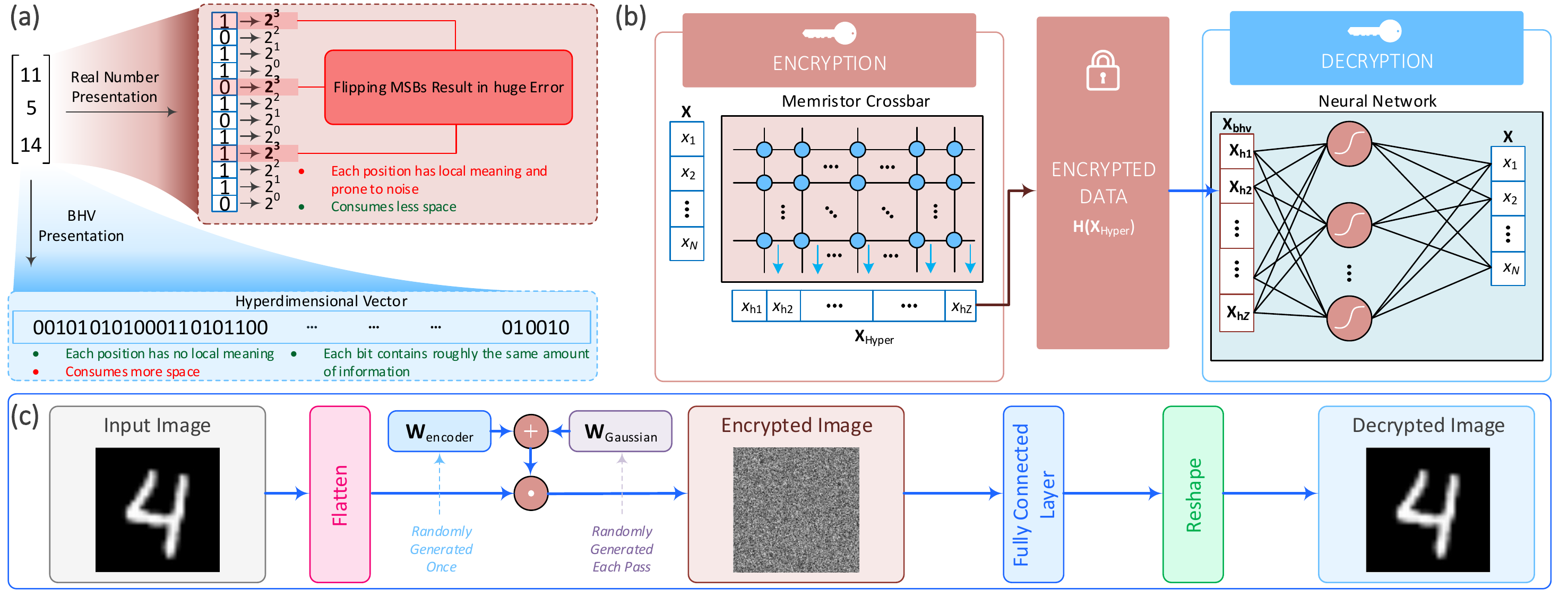}
    \caption{(a) Binary hypervector representation versus real vector representation. (b)Proposed model architecture schematic for in-memory hyperdimensional encryption. (c) Image encryption and decryption implementation.}
    \label{fig:2}
     \vspace{-6mm}
   \end{figure*}
\vspace{-1.5mm}
\section{Preliminaries}
\vspace{-0.5mm}
\subsection{Memristor Crossbar Arrays}
A typical structure of a crossbar is shown in Fig. 1, where memristors are programmed and sandwiched between the word-lines and the bit-lines. When analog voltage vectors are applied through the word-lines, memristors act as multiplication units according to Ohm’s law (i.e. $I = VG$), and the current output from the memristors are then accumulated through the bit-lines. Despite its ideal functionalities, real implementation of the crossbar often comes with non-ideality \cite{Yi2016, Ambrogio2018, Du2017, Lv2021} that causes errors in the output current vectors.
\subsection{Binary Hypervectors}
Binary hypervectors (BHV), first introduced by Kanerva \cite{Kanerva2009} in his model of hyperdimensional (HD) computing, are binary vectors with dimensions in the orders of thousands. Unlike traditional real-valued vectors that are optimized for space, BHVs are optimized for robustness. In BHV, information is distributed evenly across all entries. Such representation provides resilience against noise, non-ideality, and low resolution in computer hardware as randomly flipping one bit of information has almost no impact on the vector’s representation \cite{Kanerva2009}. Since the information is evenly spread, BHV can be more robust when its dimension is increased. Real valued vectors, on the other hand, are prone to these noises, as failures at critical bits can change the vector's representation significantly. Fig. 2(a) contrasts the two vectors.
\section{Proposed Model Architecture}
Our proposed architecture (Fig. 2(b)) consists of a hyperdimensional stochastic encoder that encrypts the ordinary real value vectors into binary hypervectors, and a multi-layer perceptron (MLP) decoder that reconstructs the original vector. 
\subsection{Hyperdimensional Stochastic Encoder}
The stochastic encoder is characterized by Equation (1), where $W_{encoder}$ is a tall, randomly initialized matrix that linearly transforms the original low dimensional input vector $\mathbf{x}$ into a \emph{hypervector},  $f_{noise}(t, W_{encoder}, \mathbf{x})$ is some noise function that depends on time $t$, $W_{encoder}$, and $\mathbf{x}$, and $\mathbf{H}$ is a binarization function defined by Equation (2), where $\epsilon$ is a hyperparameter. The result, $\mathbf{x}_{bhv}$, is an encrypted binary hypervector.\\
\begin{equation}
    \mathbf{x}_{bhv} = \mathbf{H}(W_{encoder}\mathbf{x} + f_{noise}(t, W_{encoder}, \mathbf{x}))
\end{equation}
\begin{equation}
    \mathbf{H}(x) =  \begin{cases} 
      0, \hspace{3pt} x < \epsilon \\
      1, \hspace{3pt} x > \epsilon \\
    \end{cases}
\end{equation} \par
The above formulation models the VMM of an intrinsic memristor crossbar array followed by a threshold. The randomly initialized matrix $W_{encoder}$ can be thought of as an untuned memristor crossbar, and the noise function $f_{noise}(t, W_{encoder}, \mathbf{x})$ models the crossbar non-idealities which depend on the crossbar conductance (finite conductance states and conductance variability), input voltage vectors (e.g. sneak path current), and time (cycle to cycle variability).\par
The intuition behind a hyperdimensional stochastic encoder is that the VMM operation will create hypervectors, evenly distributes the input vector's information across all entries. As a result, information at each entry can be represented by binary states. On the other hand, the impact on performance from noise reduces as the dimension of the BHV gets larger. By altering the output dimension of the stochastic encoder, we control the level of noise present in the encrypted BHV.
\vspace{-1.5mm}
\subsection{Neural Network Decoder}
\vspace{-0.5mm}
Encrypted BHV is fed into a fully connected neural network, dimension reduced by the weighted edges to reconstruct the original input vector. Weights of the neural network can be obtained through supervised learning. 
\begin{figure*}[t]
    \centering
    \includegraphics[width=\linewidth]{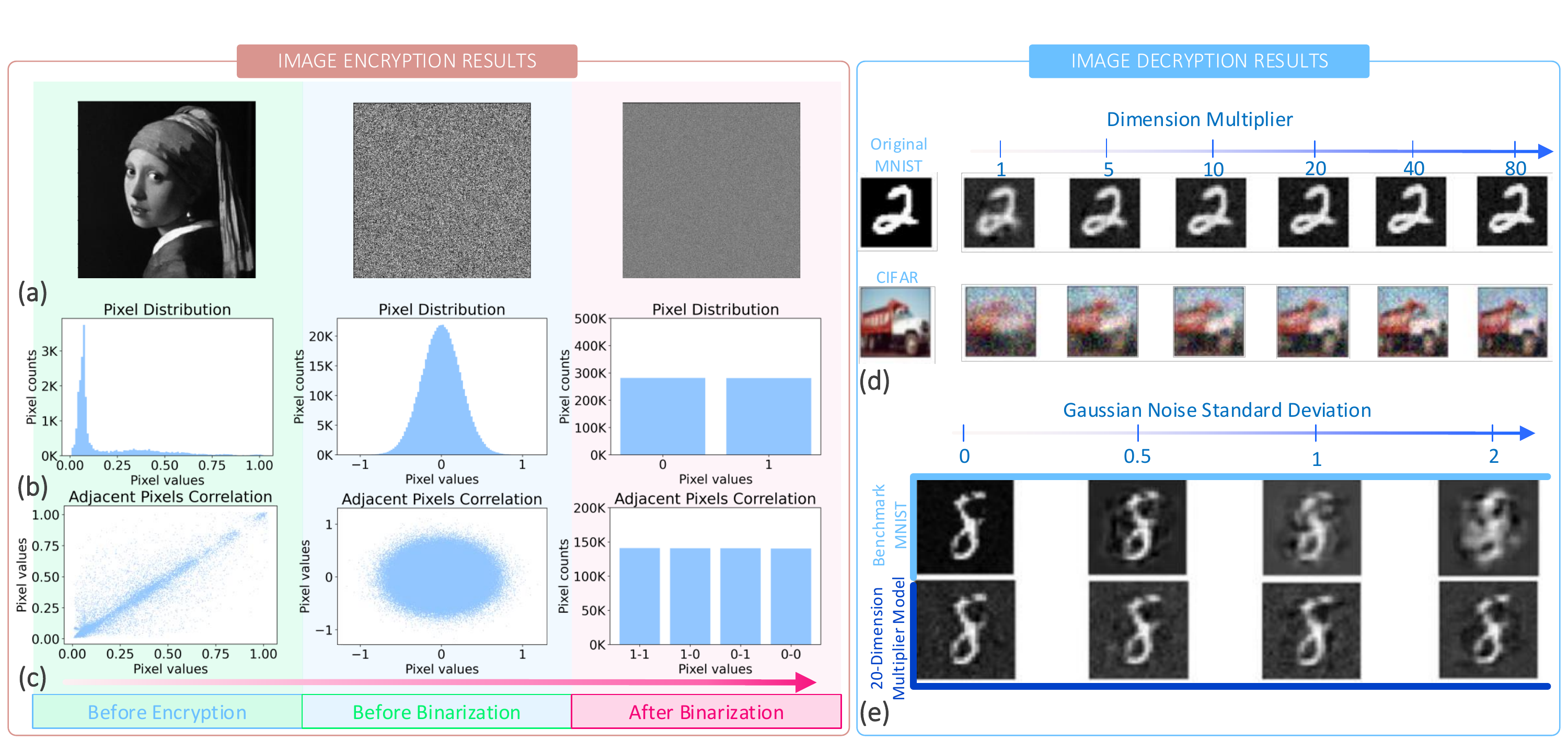}
    \caption{Image encryption and decryption results. Image encryption result is on a $150 \times 150$ pixels \textit{Girl with a Pearl Earring} image. (a) Raw image at three stages of encryption: before, after dimension expansion, and after threshold. (b) Pixel correlation at the three stages. (c) Adjacent pixel correlation at the three stages. Note that pixel correlation at the third stage is represented by a bar graph due to there being only 4 adjacent pixel combinations, namely: $0-0$, $0-1$, $1-0$, $1-1$. (d) Image decryption results: Reconstructed images at different levels of \textit{dimension multiplier} (the factor in which the input image vector dimension is expanded) in the stochastic encoding. (e) Reconstructed MNIST images at different standard deviations of Gaussian noise.}
    \label{fig:3}
     \vspace{-6.4mm}
\end{figure*}
\vspace{-0.2mm}
\section{Image Encryption}
\vspace{-0.9mm}
We first apply our proposed model for image en/decryption (Fig. 2(c)) to verify the assumptions we had in our formulation. We create a naive simulation on PyTorch, where the stochastic encoder is implemented by Equation (3).
\begin{equation}
    \mathbf{x}_{bhv} = \mathbf{H}((W_{encoder}+W_{Gaussian})\mathbf{x} )
\end{equation} \par
$W_{encoder}$ is a fixed matrix randomly generated by PyTorch's uniform initialization function in the interval $(-2,2)$ and $W_{Gaussian}$ is a Gaussian noise matrix generated by PyTorch's normal distribution function at each pass. The input vector $\mathbf{x}$ is the flattened image vector. After encoded by Equation (3), the encrypted BHV is fed into a single layer MLP to reconstruct the original image. During training, we used root mean square error as the cost function and stochastic gradient descent (SGD) to train the model until the validation loss stabilizes. We present the en/decryption result in Fig. 3.
\vspace{-1.5mm}
\subsection{Dataset and Benchmark model}
\vspace{-0.5mm}
We trained the model on MNIST and CIFAR-10 image classification datasets respectively. MNIST consists of $60$K $28\times28$ black and white images of handwritten digits from $0-9$, and CIFAR-10 consists of 60K $32\times32$ color images representing 10 classes of objects. Instead of performing classification, we trained the MLP to reconstruct these images. \par

We trained a benchmark model where $W_{encoder}$ in Equation (3) is a square matrix that does not expand the input vector $\mathbf{x}$ to hyperdimension, and we got rid of the threshold function. Ideally, the MLP decoder will learn the inverse of $W_{encoder}$ when noise is not presented, achieving little to no image reconstruction noise for decryption in a perfect scenario.
\vspace{-1.5mm}
\subsection{Simulation Results}
\vspace{-0.5mm}
We show, in Fig. 3(a-c), the raw encryption results along with the change in pixel distribution and correlation of encrypting a $150 \times 150$ pixels \textit{Girl with a Pearl Earring} image using our method. We show no obvious correlations between pixels. In Fig. 3(d), we demonstrate the positive correlation between dimension expansion and reconstructed image quality and, in Fig. 3(e), the robustness of the BHV model against noise compared to the benchmark model. Reconstructed digit still remains legible even when the Gaussian noise standard deviation is equal to $W_{encoder}$'s initialization range (-2, 2).

\begin{figure*}[t]
    \centering
    \includegraphics[width=\linewidth]{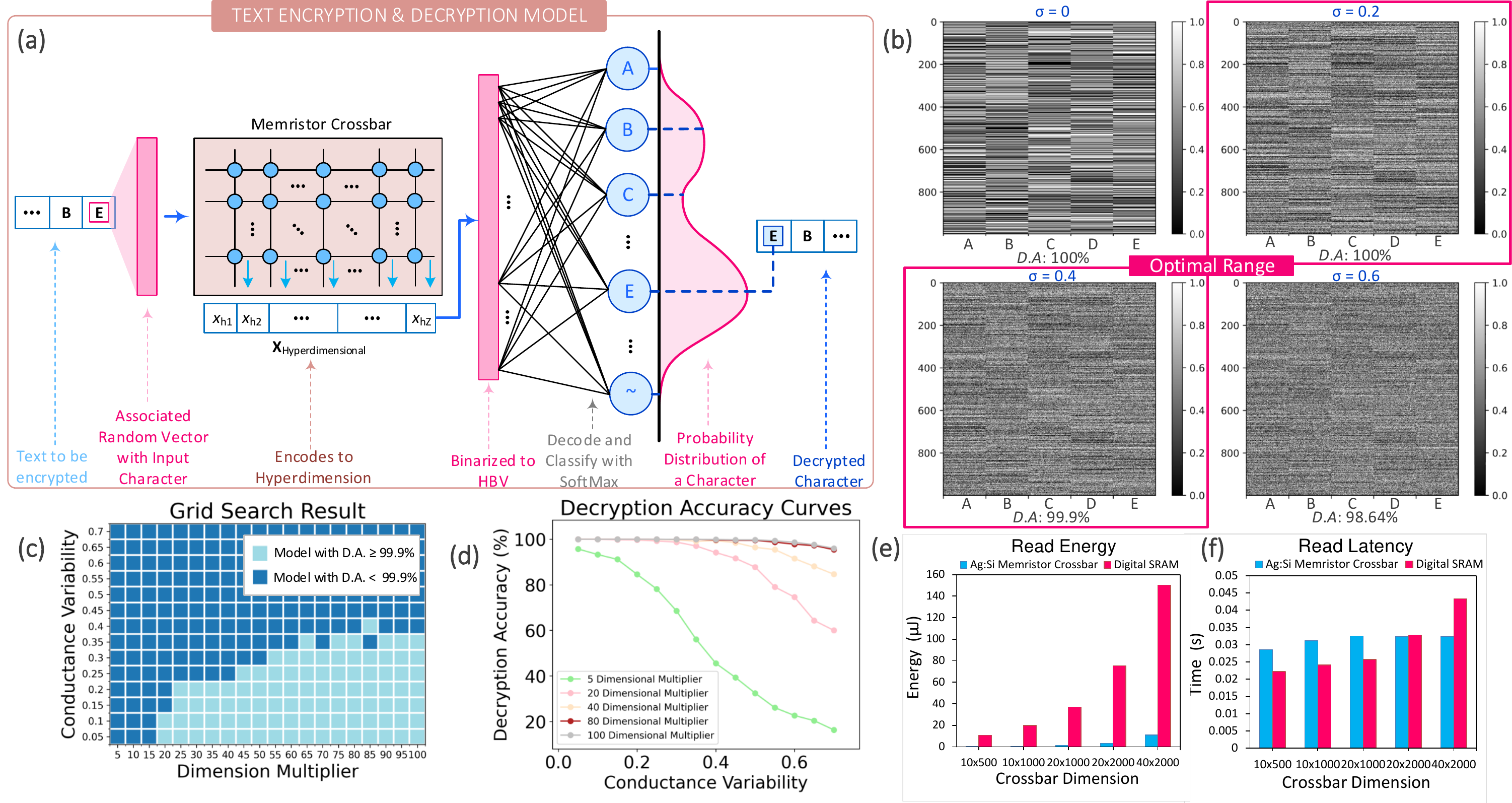}
    \caption{Application for Text Decryption and Encryption results. (a) Text en/decryption model. (b) Encrypted text by different levels of noise within the memristor crossbar. Each column is a BHV and each letter (A, B, C, D, and E) is encoded 200 times. An optimal model is achieved when both noise and decryption accuracy is high. (c) Decryption performance of models with various dimension multipliers and crossbar noise levels. Grid search on 10 dimensional secret key models with 0.05 $\textbf{P}_{on}$ and 0.05 $\textbf{P}_{off}$. ``Good" models (shaded) are those with decryption accuracy $\geq$ 99.9\% on the test set. (d) Decryption accuracy vs conductance variability for selected models. Note that more noise means more security for the encoded BHV, while less dimension multiplier means less computation cost. (e) Power and complexity analysis of memristor crossbar implementation. (f) Time and complexity analysis.}
    \label{fig:4}
    \vspace{-6mm}
\end{figure*}
\vspace{-0.5mm}
\section{Text Encryption}
\vspace{-0.5mm}
\subsection{Model Algorithm, Dataset and Simulation Setup}
In this section, we generalize our proposed model to a text en/decryption model, as shown in Fig. 4(a) and described by Algorithm 1. A new model can be generated \textit{without} new crossbar simply by regenerating secret vectors and running Train(). The dataset used in simulations are characters of $94$ classes, corresponding to ASCII 32 to 126. Our training, validation, and test sets consist of $100\mathrm{K}$, $100\mathrm{K}$, and $10\mathrm{K}$ of randomly generated characters from the 94 classes, respectively. We tested the above algorithm on a memristor crossbar array simulation based on PyTorch. Conductance variability is modeled based on Gaussian distribution and randomly stuck on/off memristors. Samples of crossbar hyperparameters are presented in Table 1 along with the test set decryption accuracy.

\begin{table}[!t]
\centering
\caption{Sample crossbar hyperparameters in experiment}
\resizebox{0.5\textwidth}{!}{
\begin{tabular}{c c c c c c c }
\toprule \toprule
\textbf{\textit{size}}& $\textbf{R}_{LRS}$(K$\Omega$)&$\textbf{R}_{HRS}$(K$\Omega$)&\textbf{$\sigma$}$^{\mathrm{1}}$&$\textbf{P}_{on}$ &$\textbf{P}_{off}$ & \textbf{D.A.$^{\mathrm{2}}$} \\
\midrule
$5\times250$& 1 & 100 & 0.1 & 0.01 & 0.01 & 99.55\% \\

$5\times500$& 1 &  100 & 0.1 & 0.01 & 0.01 & 99.96\%\\

$10\times500$& 1 &  10 & 0.1 & 0.02 & 0.02 & 100.0\%\\

$10\times1000$& 1 &  10 & 0.4 & 0.05 & 0.05 & 99.97\%\\

$15\times300$& 1 &  10 & 0.2 & 0.02 & 0.02 & 100.0\%\\

$15\times600$& 1 &  10 & 0.7 & 0.02 & 0.02 & 98.17\%\\
\bottomrule \bottomrule
\multicolumn{7}{l}{$^{\mathrm{1}}$Each conductance is perturbed by Gaussian noise with standard devia-} \\
\multicolumn{7}{l}{tion equal to $\sigma$ multiplied by the crossbar conductance range.} \\
\multicolumn{7}{l}{$^{\mathrm{2}}$Decryption accuracy over 10K characters.} \\
\end{tabular} }
\label{tab1}
\vspace{-2mm}
\end{table}

\algnewcommand{\LeftComment}[1]{\Statex \(\triangleright\) #1}
\begin{algorithm}[htb]
  \caption{Text En/Decryption}
  \begin{algorithmic}
    \State \textbf{Input: $text$, $char\_list$, $train\_set$} 
    \State {\textbf{Def }Generate($char\_list$):} \Comment{Returns a randomly generated low-dimensional secret vector \textbf{for} each char in $char\_list$.}
    \State {\textbf{Def }Map($S$, $SV$):} \Comment{Maps a String to its Secret Vector.}
    \State {\textbf{Def }VMM($\mathbf{A}$,$\mathbf{x}$):} \Comment{Performs $\mathbf{A}\mathbf{x}$ using the \textbf{crossbar}.}
    \State {\textbf{Def }Train():} \Comment{Returns a NN $Model$ (linear mapping + SoftMax) for decryption, trained by negative log likelihood loss and SGD over examples in $train\_set$.}
    \Procedure{Text En/Decryption}{ }
        \State{$Model$ = Train()} \Comment{Give $Model$ to receiver.}
        \State{$secret\_keys$ = Generate($char\_list$)}
        \For{$char$ in $text$}
        \State{$\mathbf{x}_{hyper}$ = VMM(Map($char$, $secret\_keys$))}
        \State{$\mathbf{x}_{hbv}$ = $\mathbf{H}(\mathbf{x}_{hyper})$} \Comment{This is encryption.}
        \State{... send over to receiver ...}
        \State{$c\_decrypted$ = $Model(\mathbf{x}_{hbv})$} \Comment{This is decryption.}
        \EndFor
    \EndProcedure
    \vspace{-1mm}
  \end{algorithmic}
\end{algorithm}

\vspace{-2mm}
\subsection{Experimental Results}
\vspace{-0.5mm}
We show the encrypted letters at different levels of crossbar noise in Fig. 4(b), with their corresponding decryption accuracy. Non-ideality within the memristor crossbar array makes the encoded BHV different at each pass, thereby ensuring the security of the encrypted text. In Fig. 4(c), we show the decryption performance on a crossbar with, $0.05$ $\textbf{P}_{on}$ and $0.05$ $\textbf{P}_{off}$, at various levels of dimension multipliers and conductance variability. A model is considered ``good" if the result decryption accuracy is at least 99.9\% on the test set. In Fig. 4(d), we show the decryption accuracy curves of various models. We simulated energy and time complexity on analog Ag:Si memristor crossbar and digital SRAM \cite{Chen2019} at different crossbar sizes for the stochastic encoder with NeuroSim MLP \cite{Chen2018}. Fig 4(e-f) compares the result. The Ag:Si memristor consumes $~30\times$ less energy than digital SRAM implementation while having consistent read latency at increasing the VMM scale. In addition, the digital implementation does not have the security advantage from crossbar non-ideality.
\vspace{-1.5mm}
\section{Conclusion}
Overall, we demonstrate a novel cryptography algorithm designed specifically for memristor crossbar. In the image encryption experiment, we verified our hypothesis using binary hypervectors to control crossbar noise levels. We then developed a stochastic text encryption system and demonstrated near $100$\% decryption accuracy in the text decryption with selected crossbar models. This work is a proof of concept of how memristor crossbars with their non-ideal nature can be used to directly encrypt data, paving the foundations for future works in this direction.

\bibliographystyle{IEEEtran}
\bibliography{References}

\vspace{12pt}

\end{document}